\documentclass[aps,prc,preprint,superscriptaddress,showpacs,showkeys]{revtex4-1}
\usepackage{graphicx}
\usepackage[usenames,dvipsnames,svgnames,table]{xcolor}

\begin{document}

\newcommand{\Jpsi}{J/\psi}
\newcommand{\pT}{p_{T}}
\title{{\Large Quarkonia suppression in Pb+Pb collisions at $\sqrt{s_{_{NN}}}$ =  2.76 TeV}} 
\author{\large Vineet Kumar}
\author{\large Prashant Shukla}
\email{pshukla@barc.gov.in}
\affiliation{Nuclear Physics Division, Bhabha Atomic Research Center, Mumbai, India}
\affiliation{Homi Bhabha National Institute, Anushakti Nagar, Mumbai, India}
\author{\large Ramona Vogt}
\affiliation{Nuclear and Chemical Sciences Division, Lawrence Livermore National Laboratory, Livermore, CA 94551, USA}
\affiliation{Physics Department, University of California, Davis, CA 95616, USA}

\date{\today}

\begin{abstract}
  We estimate the modification of quarkonia yields due to different processes 
  in the medium produced in Pb+Pb collisions at LHC energy. The quarkonia and heavy flavour 
  cross sections calculated up to next-to-leading order (NLO) are used in the study. Shadowing 
  corrections are obtained with the NLO EPS09 parametrization.
  A kinetic model is employed which incorporates quarkonia suppression inside a QGP, suppression 
  due to hadronic comovers, and regeneration from charm pairs. The quarkonia dissociation cross 
  section due to gluon collisions has been considered and
  the regeneration rate has been obtained using the principle of detailed balance.
  The modification in quarkonia yields due to collisions with hadronic comovers has been estimated 
  assuming that the comovers are pions.  
  The manifestations of these effects  on the nuclear modification factors for both $\Jpsi$ 
  and $\Upsilon$ in different kinematic regions  has been demonstrated for Pb+Pb collisions 
  at $\sqrt{s_{_{NN}}}$ =  2.76 TeV in comparison with the measurements.
  Both the suppression and regeneration due to a deconfined medium strongly affect 
  the low and intermediate $\pT$ range. The large observed suppression of $\Jpsi$ at $p_T~>$ 
  10 GeV/$c$ exceeds the estimates of suppression by gluon dissociation.
\end{abstract}

\pacs{12.38.Mh, 24.85.+p, 25.75.-q}
\keywords{quark-gluon plasma, quarkonia, suppression, regeneration}

\maketitle

\section{Introduction}
Heavy-ion collisions at relativistic energies are performed to create and characterize 
quark gluon plasma (QGP), a phase of strongly-interacting matter at high energy density 
where quarks and gluons are no longer bound within hadrons.
The quarkonia states ($\Jpsi$ and $\Upsilon$) have been some of the most popular tools 
since their suppression was proposed as a signal of QGP formation \cite{Matsui:1986dk}.
The understanding of these probes has evolved substantially via measurements 
through three generations of experiments: the SPS (at CERN), RHIC (at BNL) and the LHC (at CERN) 
and by a great deal of theoretical activity. (For recent reviews see 
Refs.~\cite{Schukraft:2013wba,Kluberg:2009wc,Brambilla:2010cs}.)
Quarkonia are produced early in the heavy-ion collisions and, if they evolve
through the deconfined medium, their yields should be suppressed in comparison with those in $pp$ collisions. 
The first such measurement was the `anomalous' $\Jpsi$ suppression discovered at the SPS 
which was considered to be a hint of QGP formation. The RHIC measurements showed almost the 
same suppression at a much higher energy contrary to expectation \cite{Brambilla:2010cs,Adare:2011yf}. 
Such an observation was consistent with the scenario that, at higher collision energies, the 
expected greater suppression is compensated by  $\Jpsi$ regeneration through recombination of two 
independently-produced charm quarks~\cite{Andronic:2003zv}. Since the LHC first performed Pb+Pb collisions 
at $\sqrt{s_{_{NN}}} = 2.76$ TeV, a wealth of quarkonia results have become available~\cite{Muller:2012zq,P.ShuklaforCMS:2014vna}. 
The CMS experiment carries out $\Jpsi$ measurements at high transverse momentum 
($p_T>6.5$ GeV/$c$). The nuclear modification factor $R_{AA}$ of these high $p_T$ 
prompt $\Jpsi$ decreases with increasing centrality \cite{Chatrchyan:2012np,Mironov:2013jaa} showing
moderate suppression even in the most peripheral collisions. 
Moreover $R_{AA}$ is found to be nearly independent of $p_T$ (above 6.5 GeV$/c$), showing 
that the $\Jpsi$ remains suppressed, even at very high $p_T$, up to $\sim$ 16 GeV/$c$.
By comparing with the STAR results \cite{Tang:2011kr} at RHIC it follows that the suppression of
high $p_T~\Jpsi$ has increased with collision energy.
The ALICE $\Jpsi$ results \cite{Abelev:2013ila} cover low $p_T$ and 
have little or no centrality dependence. The ALICE $\Jpsi$ suppression decreases 
substantially with decreasing $p_T$.
When compared with the PHENIX forward rapidity measurement at 
RHIC \cite{Adare:2011yf}, it suggests that low $p_T\,\Jpsi$'s are less suppressed at 
the LHC. These observations suggest $\Jpsi$ regeneration at low $p_T$ by 
recombination of independently-produced charm pairs.
At LHC energies, the $\Upsilon$ states are produced with good statistics.
The CMS measurements \cite{Chatrchyan:2011pe,Chatrchyan:2012lxa} reveal that 
the higher $\Upsilon$ states are more suppressed relative to the ground state, 
a phenomenon known as sequential suppression. The ALICE
measurements~\cite{Abelev:2014nua} at forward rapidity, ($2.5 \leq y^{\Upsilon} \leq 4.0$) are 
consistent with CMS measurements at midrapidity, $|y^{\Upsilon}|\,\leq 2.4$.

Many models were developed for the modification of quarkonia due to different processes
before the LHC startup. The suppression of quarkonia in a QGP is understood in 
terms of color screening models e.g. Ref. \cite{Matsui:1986dk,Abdulsalam:2012bw} and, 
alternatively, in terms of quarkonium dissociation by collisions with gluons \cite{Bhanot:1979vb,Xu:1995eb}. 
Statistical models \cite{Andronic:2003zv,Andronic:2012dm} can estimate of the 
regeneration of quarkonia by charm quark pairs.
The inverse of the gluon dissociation process can also be used to estimate regeneration \cite{Thews:2000rj}.  
The quarkonia yields in heavy-ion collisions are modified by non-QGP effects such as
shadowing, due to the modification of the parton distribution functions inside the nucleus,
and dissociation due to hadronic or comover interaction \cite{Vogt:2010aa}.
There have been many recent calculations to explain the LHC quarkonia results using a 
combination of the above frameworks \cite{Zhao:2011cv,Emerick:2011xu} as well as viscous
hydrodynamics~\cite{Strickland:2011mw}. 

In this paper, we calculate $\Jpsi$ and $\Upsilon$ production and
suppression in a kinetic model which includes dissociation due to thermal gluons, modification 
of the yields due to shadowing and due to collisions with comovers. Regeneration by thermal 
heavy quark pairs is also taken into account. 
Our goal is to obtain the nuclear modification factor of quarkonia as a function of transverse 
momentum and collision centrality and compare it to experimental data from CMS 
and ALICE.

\section{The production rates and cold nuclear matter effects}
{\color{black}

The heavy quark production cross section are calculated to NLO in pQCD  
using the CT10 parton densities \cite{Lai:2010vv}. The mass and scale parameters used 
for open and hidden heavy flavor production are obtained by fitting the energy dependence 
of open heavy flavor production to the measured total cross sections~\cite{Nelson:2012bc,Nelson:Future}.
Those obtained for open charm are $m_c = 1.27 \pm 0.09$~GeV,
$\mu_F/m_{T\,c} = 2.10 ^{+2.55}_{-0.85}$, and $\mu_R/m_{T\, c} = 1.60 ^{+0.11}_{-0.12}$~\cite{Nelson:2012bc}. 
The botttom quark mass and scale parameters are $m_b = 4.65 \pm 0.09$ GeV,
$\mu_F/m_{T\, b} = 1.40^{+0.75}_{-0.47}$, and $\mu_R/m_{T\, b} = 1.10^{+0.26}_{-0.19}$~\cite{Nelson:Future}.
The quarkonium production cross sections are calculated in the color evaporation model with
normalizations determined from fitting the scale parameter to the shape of the energy-dependent
cross sections~\cite{Nelson:2012bc,Nelson:Future}. The resulting uncertainty bands are smaller 
than those obtained with the fiducial parameters used in Ref.~\cite{Cacciari:2005rk}.
We note that the new results are within the uncertainties of those Ref.~\cite{Cacciari:2005rk}.  
Indeed, the charm cross sections reported at the LHC agree
better with the new values of the mass and scale than the central value of $m_c = 1.5$ GeV,
$\mu_F/m_T = \mu_R/m_T = 1$. The central EPS09 NLO parameter set~\cite{Eskola:2009uj} is used to 
calculate the modifications of the parton distribution functions (nPDF) in 
Pb+Pb collisions, referred as cold nuclear matter (CNM) effects. The CNM uncertainty is 
calculated by adding the EPS09 NLO uncertainties in quadrature.}
The production cross sections for heavy flavor and quarkonia at $\sqrt{s_{_{_{NN}}}} = 2.76$ 
TeV \cite{Kumar:2012qx} are given in Table~\ref{NLOcros}.  The yields in a minimum bias 
Pb+Pb event is obtained from the per nucleon cross
section, $\sigma_{\rm PbPb}$, in Table~\ref{NLOcros}, as
\begin{eqnarray}
N = {A^2 \sigma_{\rm PbPb} \over  
\sigma_{\rm PbPb}^{\rm tot}} \, \, .
\end{eqnarray}
 At 2.76 TeV, the total Pb+Pb cross section, $\sigma_{\rm PbPb}^{\rm tot}$, 
is 7.65 b \cite{Chatrchyan:2011sx}.

\begin{table}
\caption[]{Heavy quark and quarkonia production  cross sections at
$\sqrt{s_{_{_{NN}}}}= 2.76$ TeV. The cross sections are given per nucleon pair while
$N^{\rm PbPb}$ gives the initial number of heavy quark pair/quarkonia per Pb+Pb event.}
\label{NLOcros}
\begin{tabular}{l|l|l|l|l} 
\hline 
\hline
             & $ c \overline c$            &$\Jpsi$                      & $ b \overline b$                    & $\Upsilon$   \\              
\hline
$\sigma_{pp}$ & $4.11^{+2.69}_{-2.50}$ mb    & $21.6^{+10.6}_{-10.4}~\mu$b   & $110.5^{+15.1}_{-14.2}~\mu$b            & $0.22^{+0.07}_{-0.06}~\mu$b  \\

$\sigma_{\rm PbPb}$ & $3.21^{+2.1}_{-1.95}$ mb    &16.83$^{+8.26}_{-8.10}~\mu$b    & $100.5^{+13.7}_{-12.9}~\mu$b             & 0.199$^{+0.063}_{-0.054}~\mu$b  \\

$N^{\rm PbPb}$     & $18.12^{+12}_{-11}$       & $0.0952^{+0.047}_{-0.046}$         & $0.57^{+0.08}_{-0.07}$                          & $0.001123^{+0.0004}_{-0.0003}$       \\

\hline
\hline
\end{tabular}
\end{table}
\section{Modification of quarkonia in the presence of QGP}
 In the kinetic approach \cite{Thews:2000rj}, the proper time $\tau$ evolution of the quarkonia 
population $N_{Q}$
is given by the rate equation 

\begin{equation}\label{eqkin}
{dN_{Q} \over d\tau}  =  - \lambda_D  \rho_g N_{Q} + \lambda_F {N_{q \bar{q}}^{2} \over V(\tau)},
\end{equation}
where $V(\tau)$ is the volume of the deconfined spatial region and $N_{q \bar{q}}$ is the number of initial 
heavy quark pairs produced per event depending on the centrality defined by the number of participants
$N_{\rm part}$.
 The $\lambda_{D}$ is the dissociation rate obtained by the dissociation cross section averaged over 
the momentum 
distribution of gluons and $\lambda_{F}$ is the formation rate obtained by the formation cross section 
averaged over the momentum distribution of heavy quark pair $q$ and $\bar{q}$. 
$\rho_g$ is the density of thermal gluons.
 The number of quarkonia at freeze-out time $\tau_f$ is given by the solution of Eq.~(\ref{eqkin}),
\begin{equation}
N_{Q}(p_T) = S(p_T) \,N_{Q}^{\rm PbPb}(p_T)+N_{Q}^F(p_T).
\label{eqbeta}
\end{equation}
Here $N_{Q}^{\rm PbPb}(p_T)$ is the number of initially-produced quarkonia (including shadowing)
as a function of $p_T$ and $S(p_T)$ is their survival probability from gluon collisions at freeze-out, 
\begin{equation}
S(p_T) = \exp \left( {-\int_{\tau_0}^{\tau_f}f(\tau) \lambda_{\rm D}(T,p_T)\,\rho_g(T)\,d\tau} \right).
\end{equation}
 The temperature $T(\tau)$ and the QGP fraction $f(\tau)$ evolve from initial time $\tau_0$ 
to freeze-out time $\tau_f$ due to expansion of the QGP. The initial temperature and the 
evolution is dependent on collision centrality $N_{\rm part}$.
$N_{Q}^F(p_T)$ is the number of regenerated quarkonia per event,
\begin{equation}
N_{Q}^F(p_T)=S(p_T)N_{q \bar{q}}^{2} \int_{\tau_0}^{\tau_f}{{\lambda_{\mathrm{F}}(T,p_T) \over V(\tau)\,S(\tau,p_T)} d\tau}.
\end{equation}
The nuclear modification factor ($R_{AA}$) can be written as 
\begin{equation}
R_{AA}(p_T)=S(p_T) \, R(p_T) + \frac{N_{Q}^F(p_T)}{N_{Q}^{pp}(p_T)}.
\label{raa}
\end{equation}
Here $R(p_T)$ is the shadowing factor.
$R_{AA}$ as a function of collision centrality, including regeneration, is
\begin{equation}
R_{AA}(N_{\rm part}) = \frac{\int_{p_{T\,\rm cut}} N_{Q}^{pp}(p_T)S(p_T)\, R(p_T) dp_T}{\int_{p_{T\,\rm cut}} N_{Q}^{pp}(p_T) dp_T} + 
\frac{\int_{p_{T\, \rm cut}} N_{Q}^F(p_T) dp_T}{\int_{p_{T\, \rm cut}} N_{Q}^{pp}(p_T) dp_T}
\label{raa2}
\end{equation}
Here $p_{T~{\rm cut}}$ defines the $p_T$ range for a given experimental acceptance.
 $N_{Q}^{pp}(p_T)$ is the unmodified $p_T$ distribution of quarkonia obtained by NLO 
calculations and scaled to a particular centrality of the Pb+Pb collisions.

\begin{figure}
\includegraphics[width=0.49\textwidth]{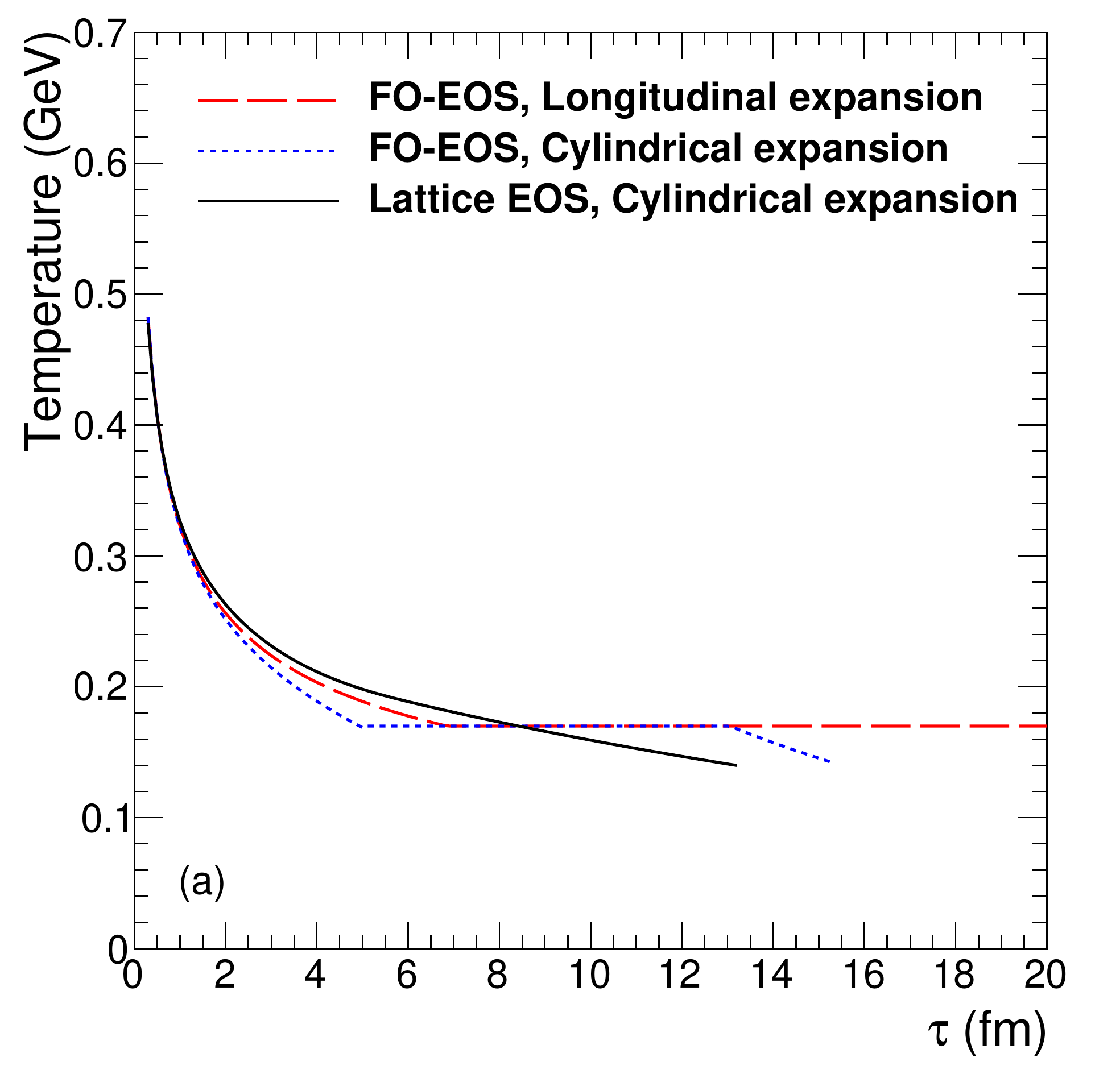}
\includegraphics[width=0.49\textwidth]{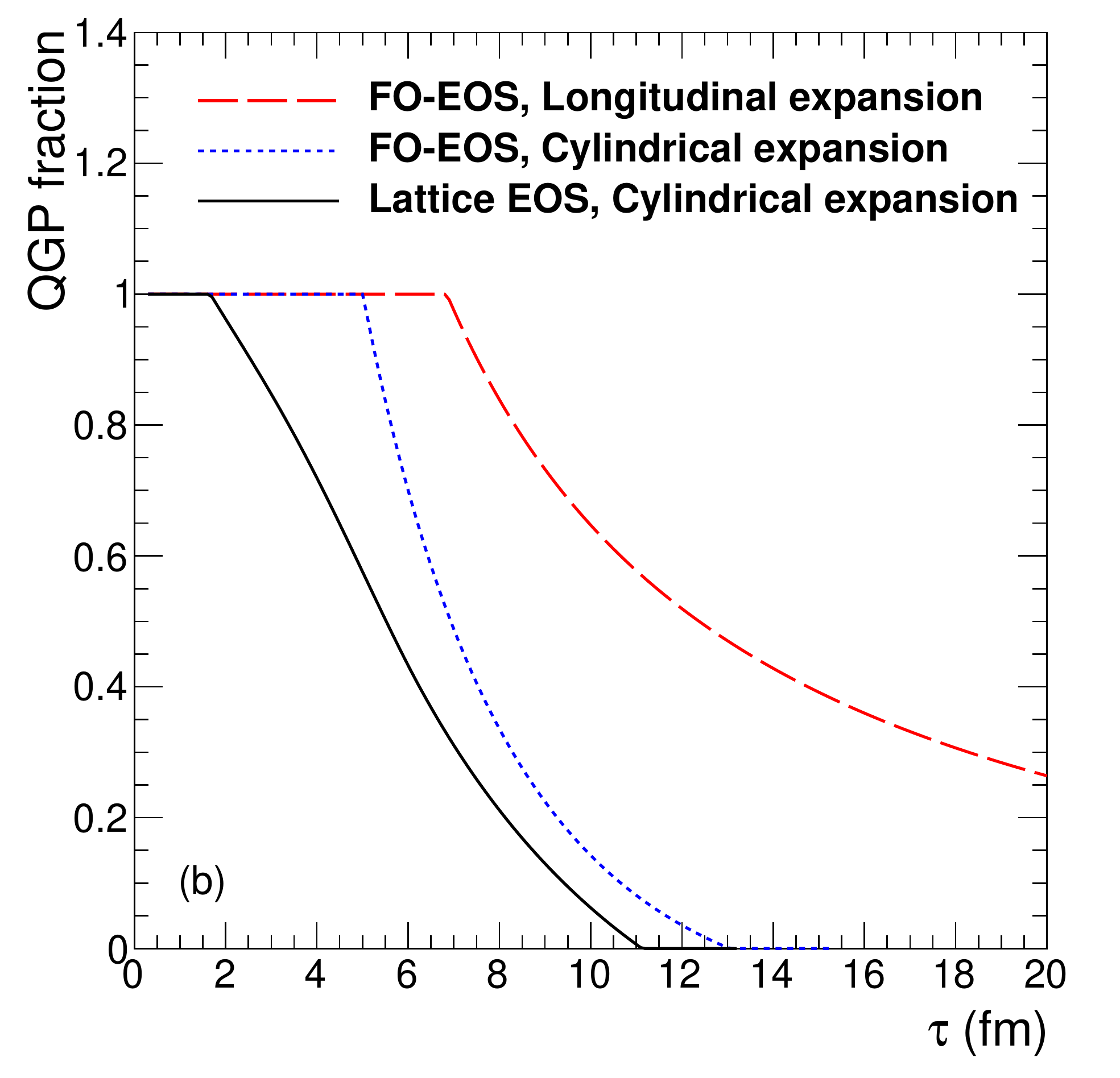}
\caption{(Color online) (a) Temperature and (b) QGP fraction in the system as a function of proper 
time $\tau$ in case of the most central (0-5$\%$) collisions for longitudinal and cylindrical expansions 
using first order and lattice equation of state . }
\label{fig:TauVsTemp}
\end{figure}

The evolution of the system for each centrality bin is governed by
an isentropic cylindrical expansion with volume element
\begin{equation}
V(\tau) = \tau\,\pi\,(R + {1\over 2} a_{T} \, \tau^2 )^{2},
\end{equation}
 where $a_{T} = 0.1\,c^2$ fm$^{-1}$ is the transverse acceleration \cite{Zhao:2011cv}.
 The initial transverse size, $R$, as a function of centrality is
\begin{equation}
R(N_{\rm part}) = R_{0-5\%} \, \sqrt{N_{\rm part} \over (N_{\rm part})_{0-5\%} },
\label{RVsNPart}
\end{equation}
where $R_{0-5\%} = 0.96\,R_{\rm Pb}$ and $R_{\rm Pb}$ is the radius of the lead nucleus.
The evolution of entropy density for each centrality is obtained by entropy conservation, 
 $s(T)\,V(\tau)= s(T_0)\,V(\tau_0)$.
The equation of state (EOS) obtained from Lattice QCD, along with a hadronic resonance gas, \cite{Huovinen:2009yb} 
is used to obtain the temperature as a function of proper time $\tau$.
  The initial entropy density for each centrality is calculated using 
\begin{equation}
s(\tau_0) = s(\tau_0)|_{0-5\%} \left(\frac{dN/d\eta}{N_{\rm part}/2}\right)\left(\frac{dN/d\eta}{N_{\rm part}/2}\right)^{-1}_{0-5\%}.
\label{TempNpart}
\end{equation}
  Measured values of $(dN/d\eta)/(N_{\rm part}/2)$ as a function of $N_{\rm part}$ 
{\color{black} \cite{Aamodt:2010cz,Chatrchyan:2011pb}} are used in the calculations.
The initial entropy density, $s(\tau_0)|_{0-5\%}$, for 0-5\% centrality is 
\begin{eqnarray}
s(\tau_0)|_{0-5\%}  = {a_{m} \over V(\tau_0)|_{0-5\%}}   \left(\frac{dN}{d\eta}\right)_{0-5\%} . 
\label{TempVsMult}
\end{eqnarray}  
Here $a_m$ {\color{black}(= 5)} is a constant which relates the total entropy to the total 
multiplicity $dN/d\eta$. It is obtained from hydrodynamic calculations \cite{Shuryak:1992wc}.
{\color{black}
We estimate the initial temperature, $T_0$, in the 0-5$\%$ most central collisions
from the total multiplictity in the rapidity region of interest, assuming that the initial time is
$\tau_0 = 0.3$ fm/$c$ over all rapidity.  The total multiplicity in a given rapidity region is
3/2 times the charged particle multiplicity in Pb+Pb collisions at 2.76 TeV.  With the lattice
EOS, at midrapidity, with $(dN_{\rm ch}/d\eta)_{0-5\%} = 1600$~\cite{Aamodt:2010cz,Chatrchyan:2011pb}, 
we find $T_0 = 0.484$ GeV.  Likewise, at forward rapiidity, $2.5 \leq y \leq 4$~\cite{Abbas:2013bpa}, 
$T_0 = 0.427$ GeV.}
The (proper) time evolution of temperature is shown in Fig.~\ref{fig:TauVsTemp}(a) 
and that of QGP fraction in Fig.~\ref{fig:TauVsTemp}(b), in the case of the most central (0-5$\%$) collisions.
Here we compare the evolution obtained with longitudinal and cylindrical expansions using 
both a first order and the lattice EOS. 
 For the first order EOS, $T_c$ = 0.170 GeV. The QGP fraction goes from 1 to 0 at $T_c$ 
assuming a mixed phase of QGP and hadrons. The QGP fraction in case of lattice EOS governs the 
number of degrees of freedom, decided by the entropy density. It is fixed to unity above an entropy density 
corresponding to a 2-flavour QGP and fixed to zero below entropy density for a hot 
resonance gas. The freeze out temperature in all cases is $T_f=0.140$ GeV.  
\subsection{Dissociation Rate}
In the color dipole approximation, the gluon dissociation cross section as function of gluon energy, $q^0$,
 in the quarkonium rest frame is~\cite{Bhanot:1979vb}
\begin{equation}
\sigma_{D}(q^{0}) = {8\pi \over 3} \, {16^2 \over 3^2} {a_0 \over m_q}  \frac{(q^0/\epsilon_0 - 1)^{3/2}} {(q^0/\epsilon_0)^5},
\end{equation}
 where $\epsilon_0$ is the quarkonia binding energy and $m_q$ is the charm/bottom quark mass 
and $a_0=1/\sqrt{m_q\epsilon_0}$.
 The values of $\epsilon_0$ are taken as 0.64 and 1.10 GeV for the ground states, $\Jpsi$ and $\Upsilon$(1S),
respectively \cite{Karsch:1987pv}.
 For the first excited state of bottomonia, $\Upsilon$(2S), we use dissociation
cross section from Ref.~\cite{Arleo:2001mp}.

\begin{figure}
\includegraphics[width=0.60\textwidth]{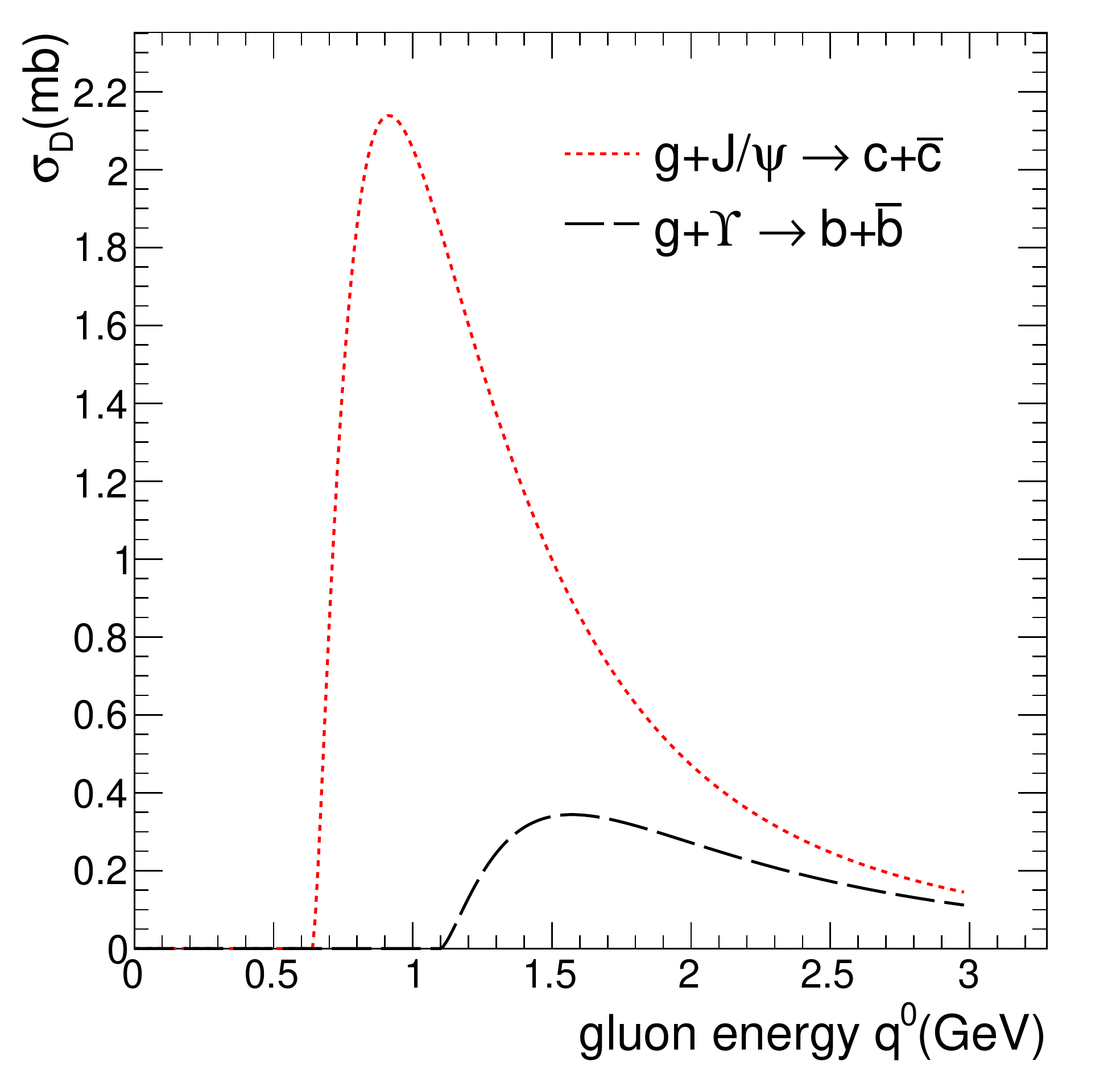}
\caption{(Color online) Gluon dissociation cross section of quarkonia as a function of gluon energy ($q^{0}$) in
quarkonia rest frame.}
\label{fig:SigmaDQ0}
\end{figure}

\begin{figure}
\includegraphics[width=0.49\textwidth]{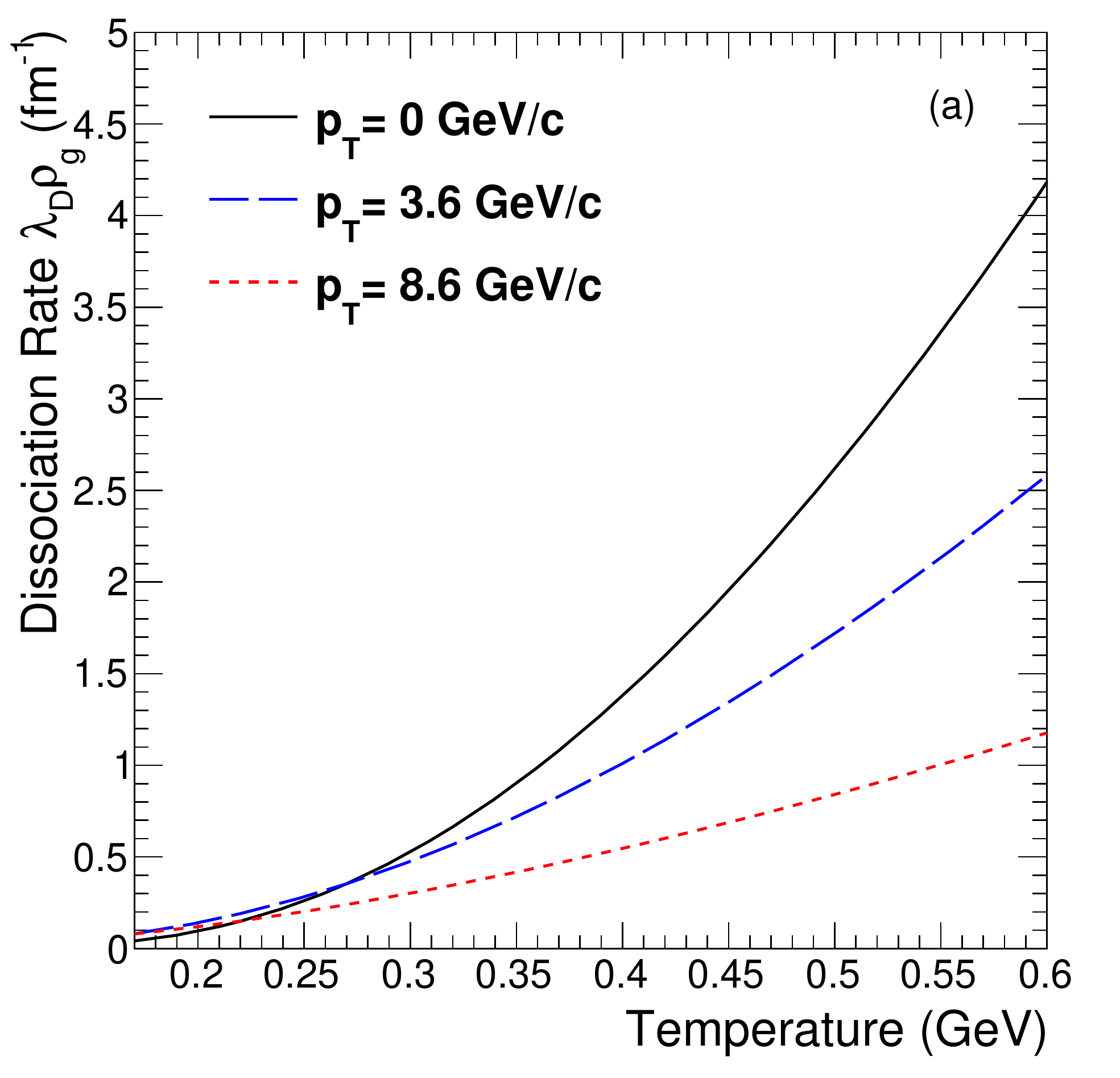}
\includegraphics[width=0.49\textwidth]{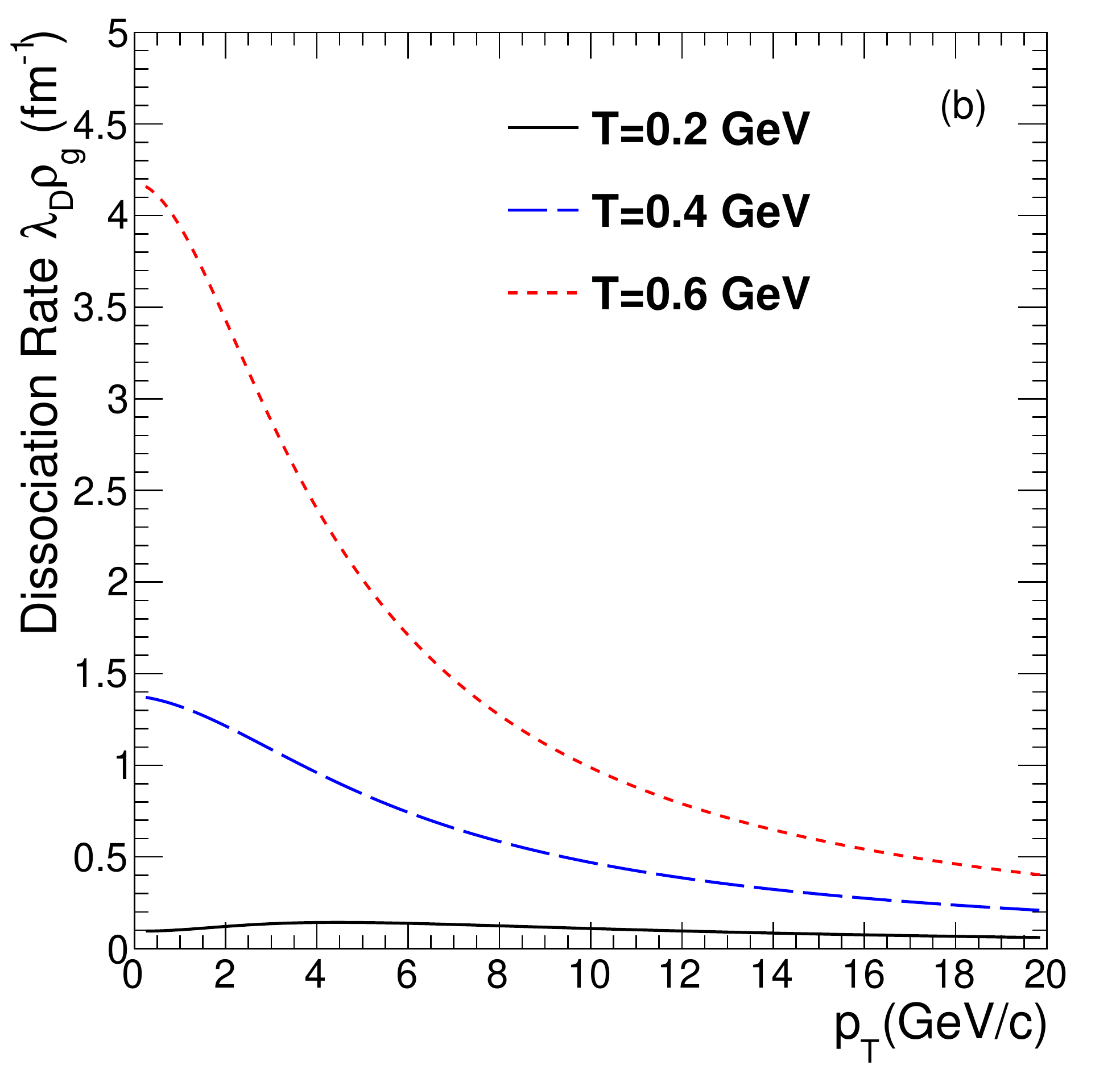}
\caption{(Color online) Gluon dissociation rate of $\Jpsi$ as a function of (a) temperature and  
(b) transverse momentum.}
\label{fig:DRateVsTempAndPt}
\end{figure}

Figure \ref{fig:SigmaDQ0} shows the gluon dissociation cross sections of $\Jpsi$ and $\Upsilon$(1S)
as a function of gluon energy. The dissociation cross section is zero when the gluon energy is less 
than the binding energy of the quarkonia. It increases with gluon energy and reaches a maximum at 1.2 (1.5) GeV for 
$\Jpsi~(\Upsilon(1{\rm S}))$. At higher gluon energies, the interaction probability decreases. The gluon energy $q^0$ 
is related to the square of the center of mass energy $s$, of the quarkonium-gluon system by
\begin{eqnarray}
 q^{0} = \frac{s-M_{Q}^{2}}{2\,M_{Q}}
\end{eqnarray}  
where $s=M_{Q}^{2} + 2  p_g \, \sqrt{M_{Q}^2 + p^2} - 2  p_g \, p \, {\rm cos\theta}$, and $M_{Q}$ and $p$ 
are mass and momentum of quarkonium and $\theta$ is angle between the quarkonium and the gluon.
We calculate the dissociation rate as a function of quarkonium momentum 
by integrating the dissociation cross section over thermal gluon momentum 
distribution $f_{g}(p_g)$,   
\begin{eqnarray}
\lambda_{D} \rho_{g}  & = & \langle \sigma v_{\rm rel} \rangle \,\rho_{g}  = \frac{g_g}{(2\pi)^{3}} \int d^{3}p_g \, f_{g}(p_g)  \, \sigma_{D}(s) v_{\rm rel}(s)  \nonumber \\ 
                   & = & \frac{g_g}{(2\pi)^{3}} \int dp_g 2\pi p_g^{2} f_{g}(p_g) \int d\,{\rm cos\theta}\,\sigma_{D}(s)\,v_{\rm rel}(s),
\end{eqnarray}
where $\sigma_{D}(s) = \sigma_{D}(q^0(s))$.
 The relative velocity, $v_{\rm rel}$, between the quarkonium and the gluon is
\begin{eqnarray}
 v_{\rm rel}  = {s- M_{Q}^{2} \over 2  p_g\sqrt{M_{Q}^2 + p^2}}.  
\label{eq7}
\end{eqnarray}
The $\Jpsi$ gluon dissociation rates as a function of $T$ are shown in 
Fig.~\ref{fig:DRateVsTempAndPt}(a) and as a function of $p_T$ in Fig.~\ref{fig:DRateVsTempAndPt}(b).
The dissociation rate increases with temperature due to the increase in gluon density. 
The dissociation rate is maximum when the quarkonium is at rest and decreases with $p_T$.
\subsection{Formation Rate}
\begin{figure}
\includegraphics[width=0.49\textwidth]{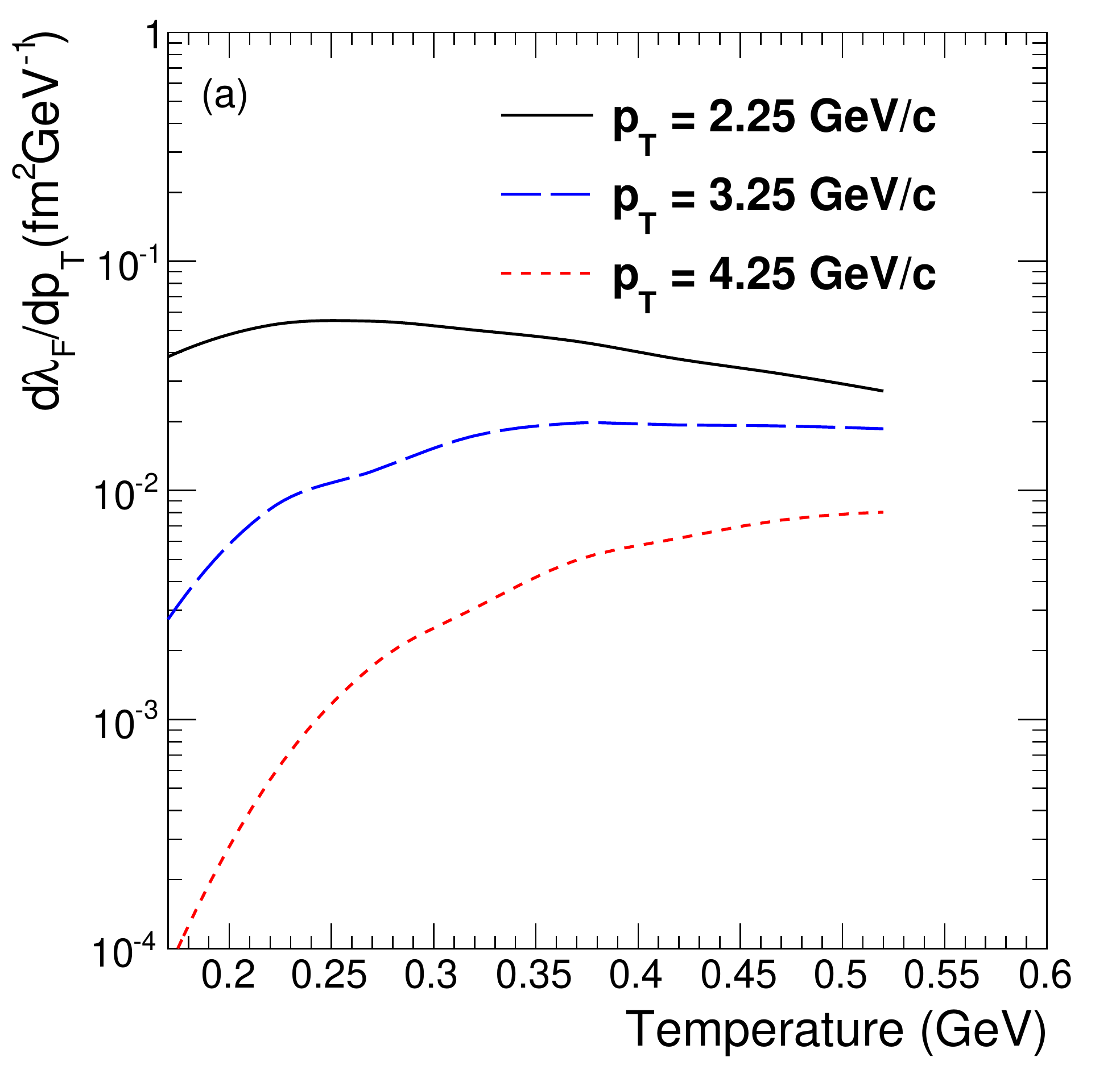}
\includegraphics[width=0.49\textwidth]{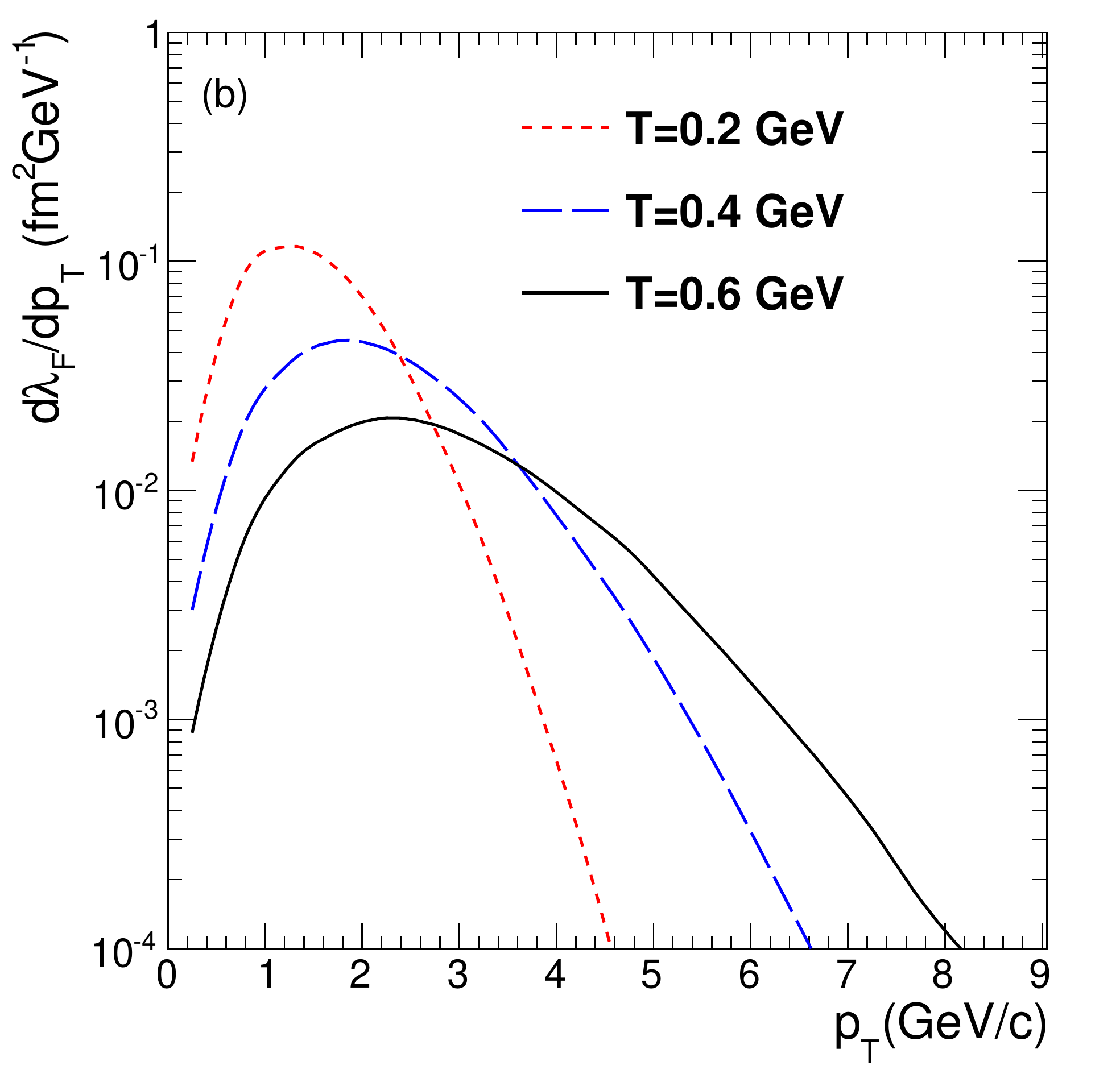}
\caption{(Color online) Formation rate of  $\Jpsi$ as a function of (a) temperature and 
(b) transverse momentum.}
\label{fig:ForRateVsTempAndPt}
\end{figure}

We can calculate the formation cross section from the dissociation cross section using detailed balance~\cite{Thews:2000rj,Thews:2005vj},
\begin{equation}
\sigma_{F} = \frac{48}{36}\,\sigma_{D}(q^0)\frac{(s-M_{Q}^2)^{2}}{s(s-4m_q^{2})}.
\end{equation}
The formation rate of quarkonium with momentum {\bf p} can be written as
\begin{equation}
\frac{d\lambda_{F}}{d{\rm\bf p}} = \int \,d^{3}p_1 \,d^{3}p_2 \,\sigma_{F}(s)\, v_{\rm rel}(s)\,f_{q}(p_1)\, f_{\bar{q}} (p_2)\,\delta({\rm\bf p}-( {\rm\bf p_1} + {\rm\bf p_2} )).
\end{equation}
Here $f_{q/\bar{q}}(p)$ are taken as thermal distribution function of  $q/\bar{q}$ which are 
normalized to one, $\int f_{q}(p) d^{3}p  = 1 $ and $v_{\rm rel}$ is relative velocity of the
$q\bar{q}$ quark pair,
\begin{eqnarray}
v_{\rm rel} &=& {\sqrt{(p_{1}.p_{2})^{2} - m_q^{4} } \over E_{1} \, E_{2}}. 
\end{eqnarray}
Here $p_1 = (E_1,{\bf p_{1}})$ and $p_{2} = (E_{2},{\bf p_{2}})$ are the four momenta of the heavy quark and 
antiquark respectively.
Figure \ref{fig:ForRateVsTempAndPt} (a) shows the variation of the formation rate as a function 
of $T$ and Fig.~\ref{fig:ForRateVsTempAndPt} (b) shows as a function of $\Jpsi$ $p_T$.
The $\Jpsi$ generated from recombination of uncorrelated heavy quark pairs will have 
softer $p_{T}$ distributions than those of $\Jpsi$'s coming from the initial hard scatterings.
Thus the effect of recombination will be important only at low $p_T$.

\begin{figure}
\includegraphics[width=0.49\textwidth]{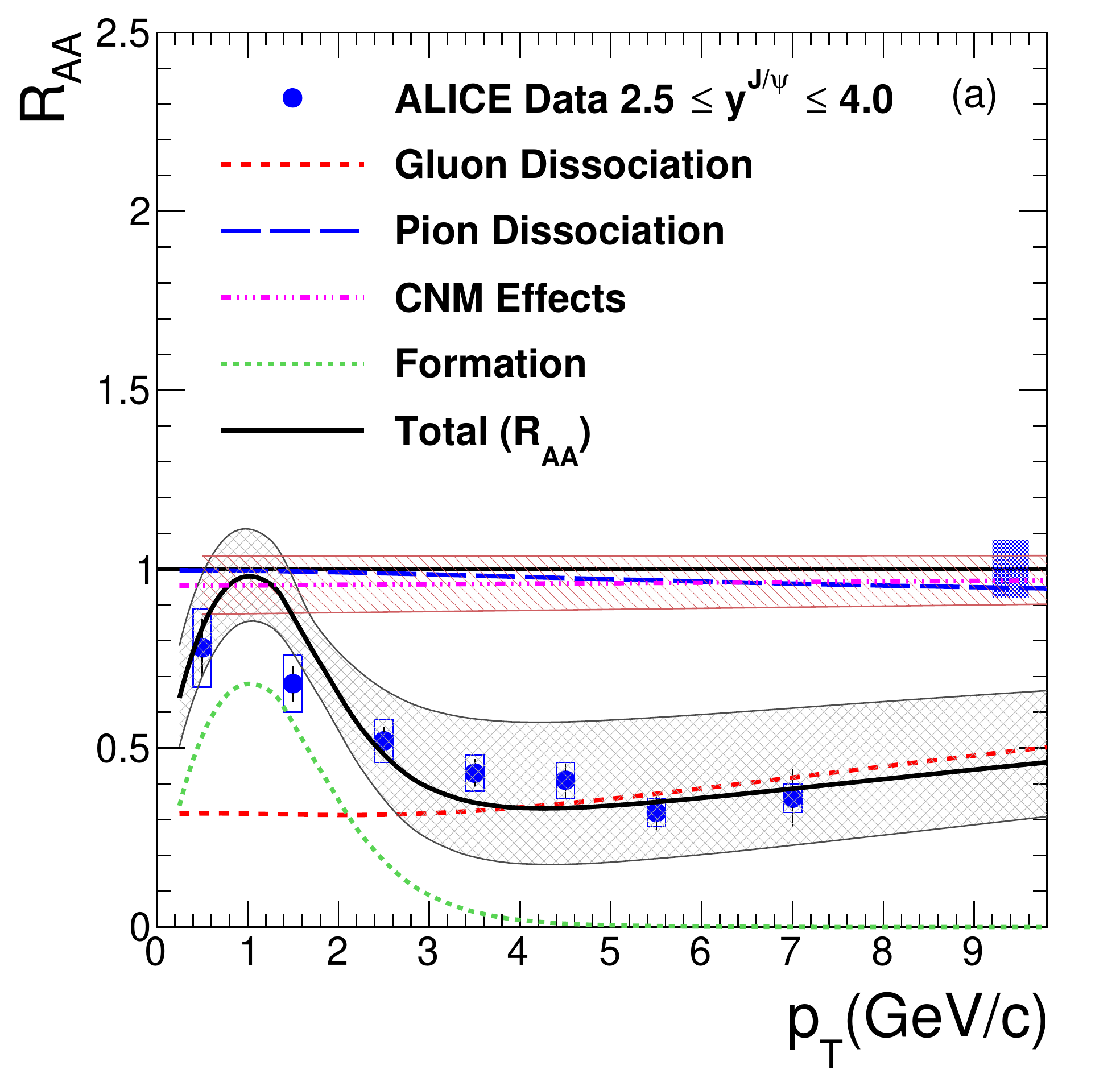}
\includegraphics[width=0.49\textwidth]{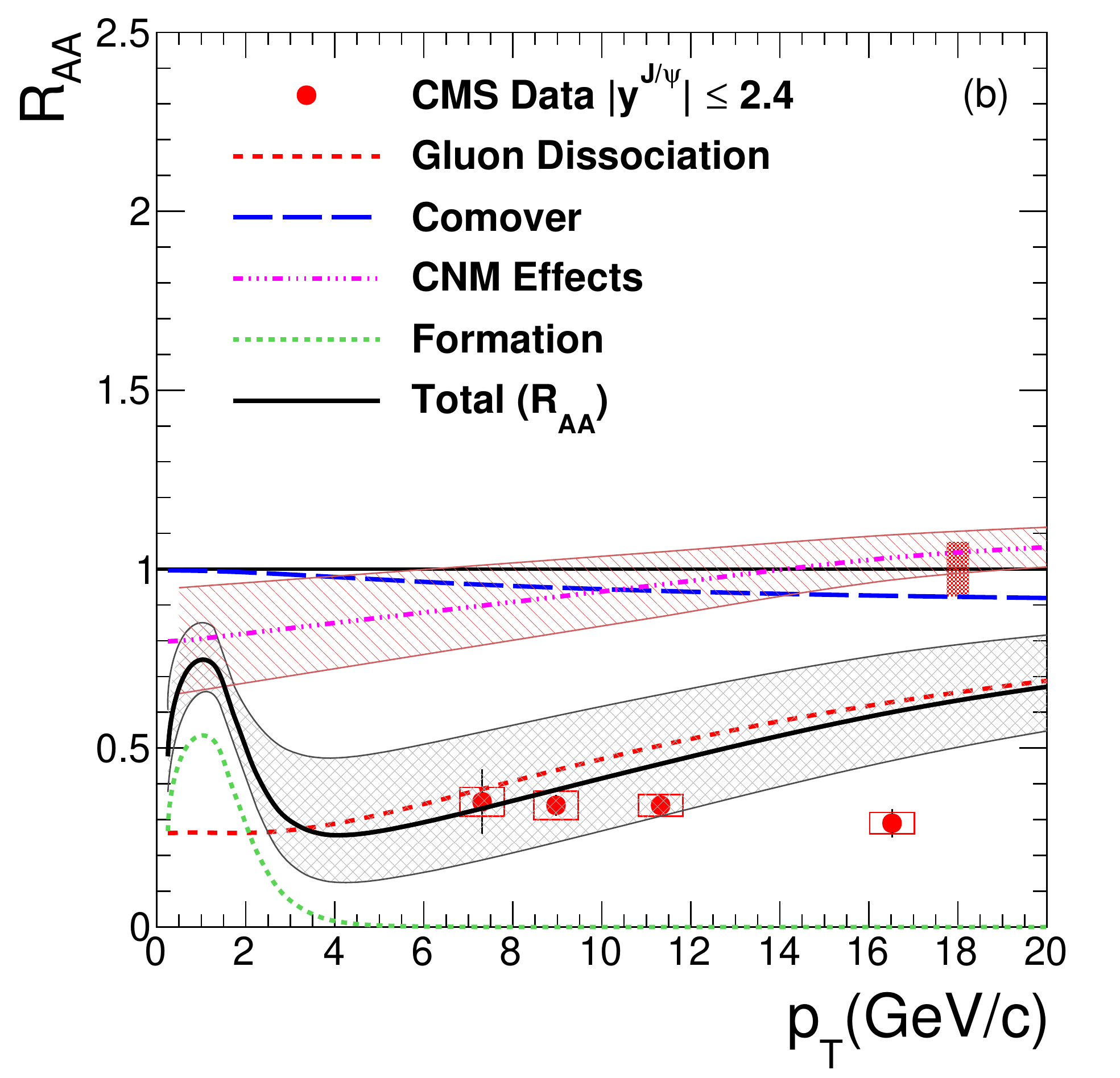}
\caption{(Color online) Calculated nuclear modification factor ($R_{AA}$) as a function of $\Jpsi$ 
transverse momentum compared with (a) ALICE and (b) CMS measurements.}
\label{fig:JPsiRaaVsPt}
\end{figure}
\begin{figure}
\includegraphics[width=0.60\textwidth]{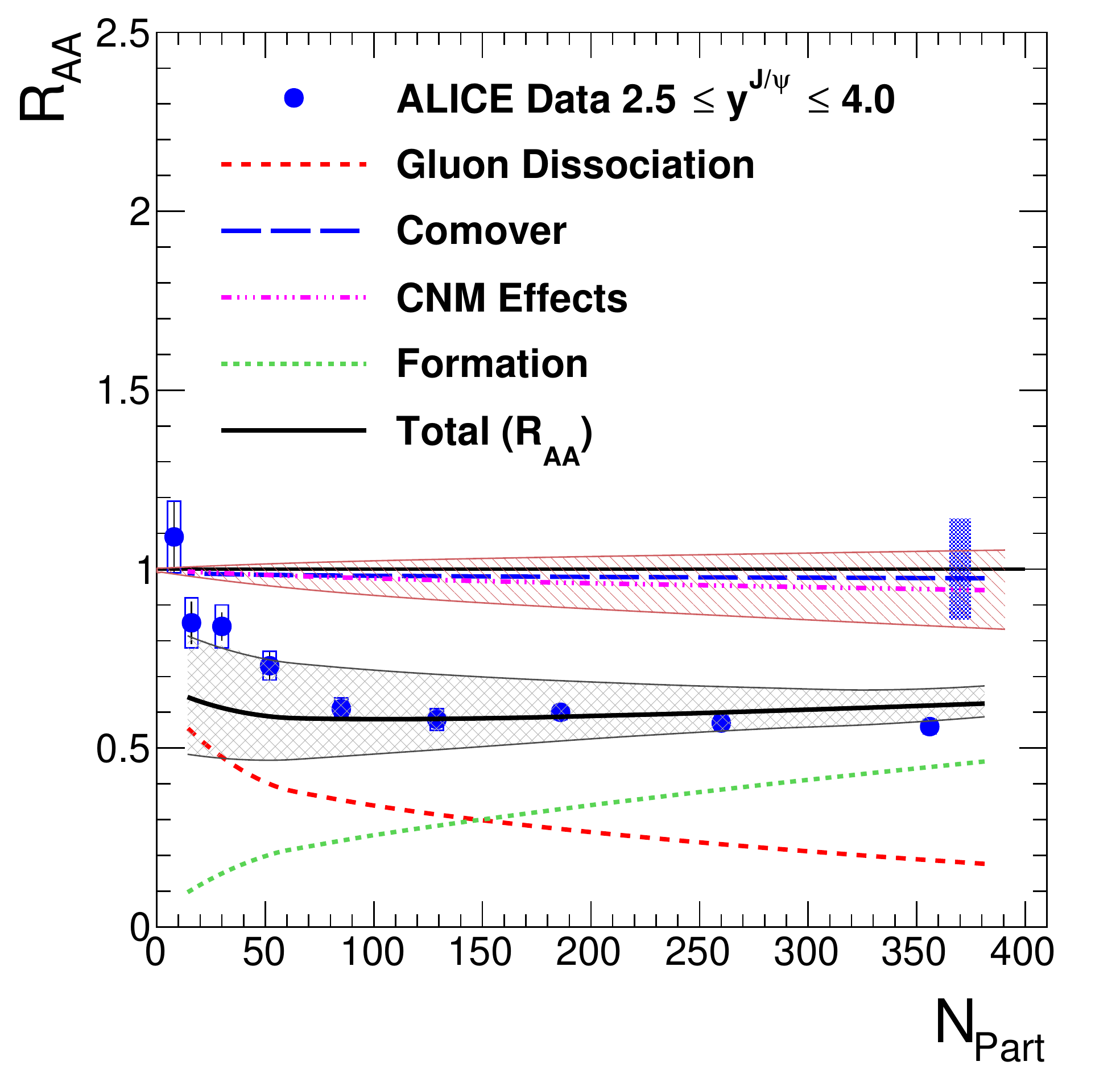}
\caption{(Color online) Calculated nuclear modification factor ($R_{AA}$) compared with ALICE 
  measurements at LHC.}
\label{fig:ALICEJPsiRaa_Forward}
\end{figure}
\begin{figure}
\includegraphics[width=0.49\textwidth]{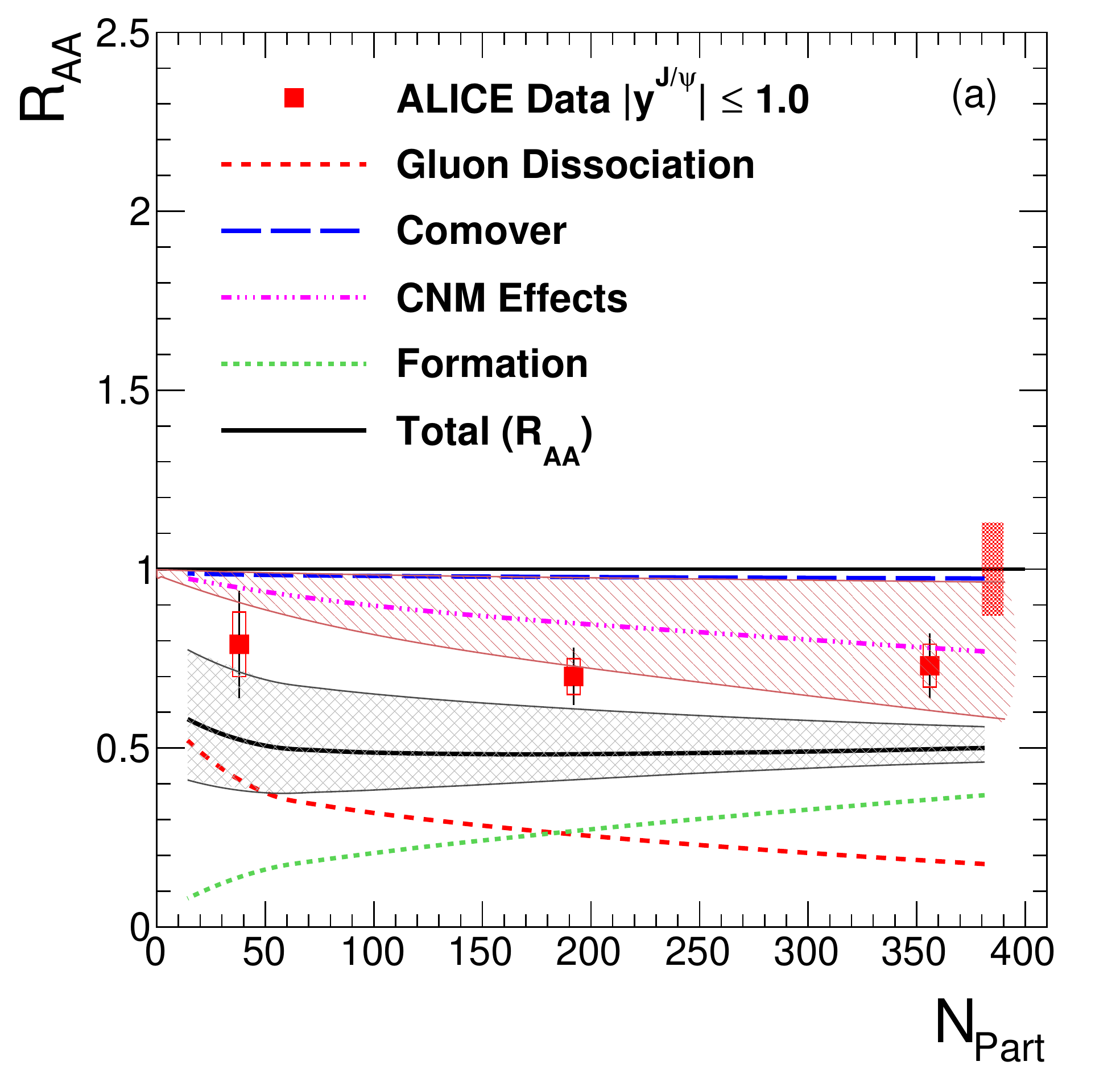}
\includegraphics[width=0.49\textwidth]{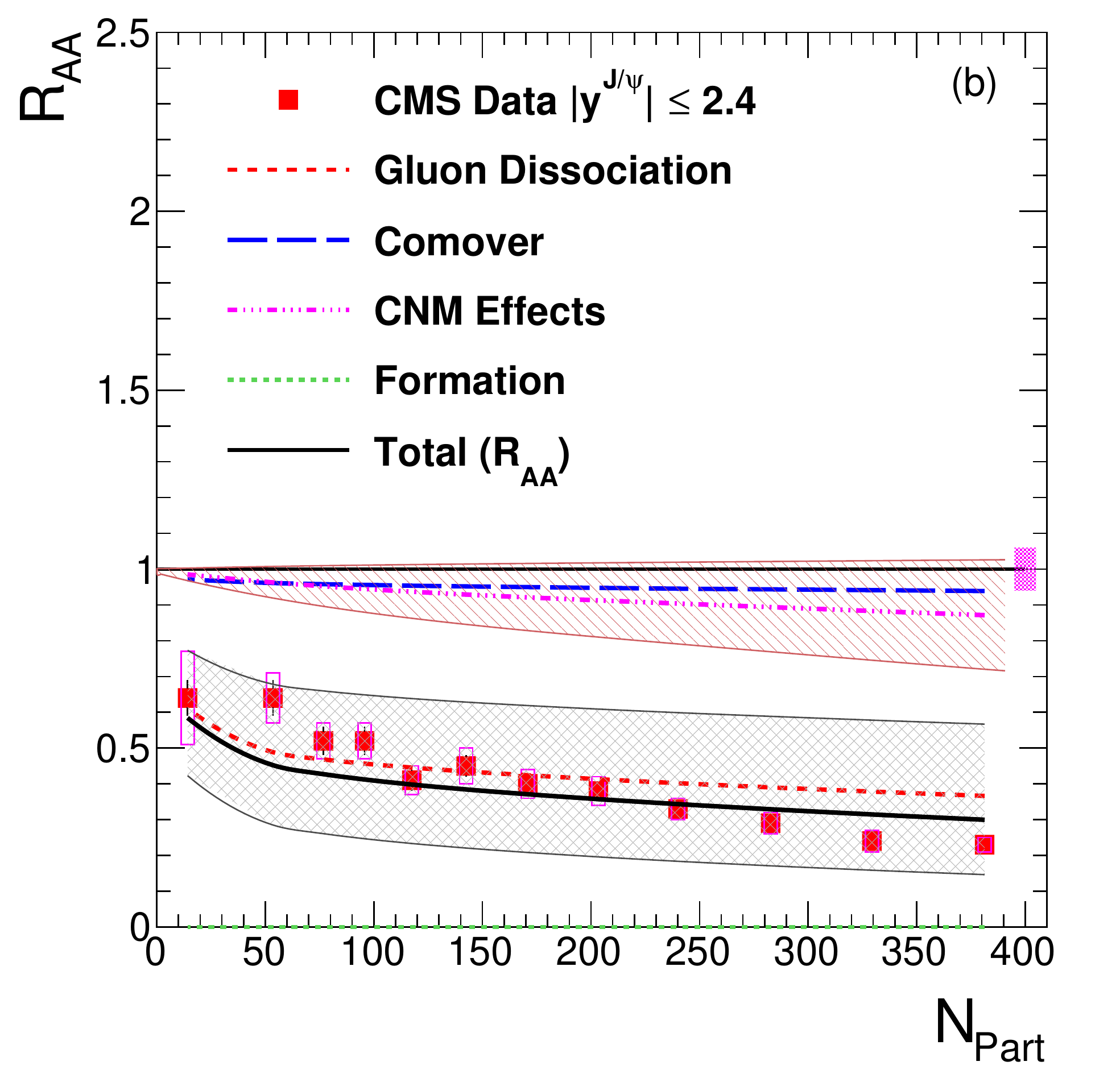}
\caption{(Color online) Calculated nuclear modification factor ($R_{AA}$) compared with (a) ALICE and 
(b) CMS measurements at midrapidity. The regeneration for the CMS high $p_T$ measurement is negligible 
in comparison to the low $p_T$ ALICE measurement.} 
\label{fig:JPsiRaa_Mid}
\end{figure}
\begin{figure}
\includegraphics[width=0.49\textwidth]{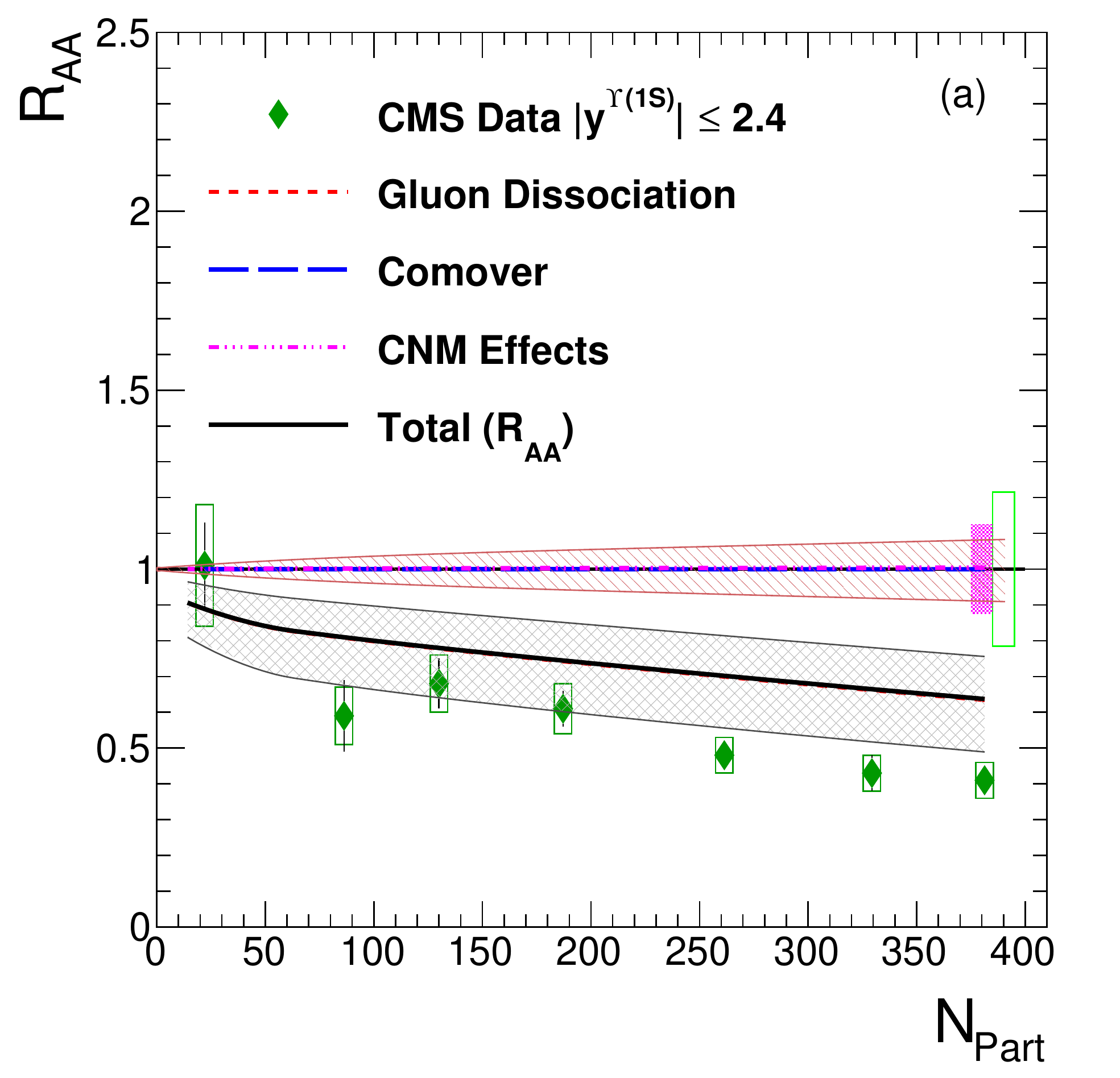}
\includegraphics[width=0.49\textwidth]{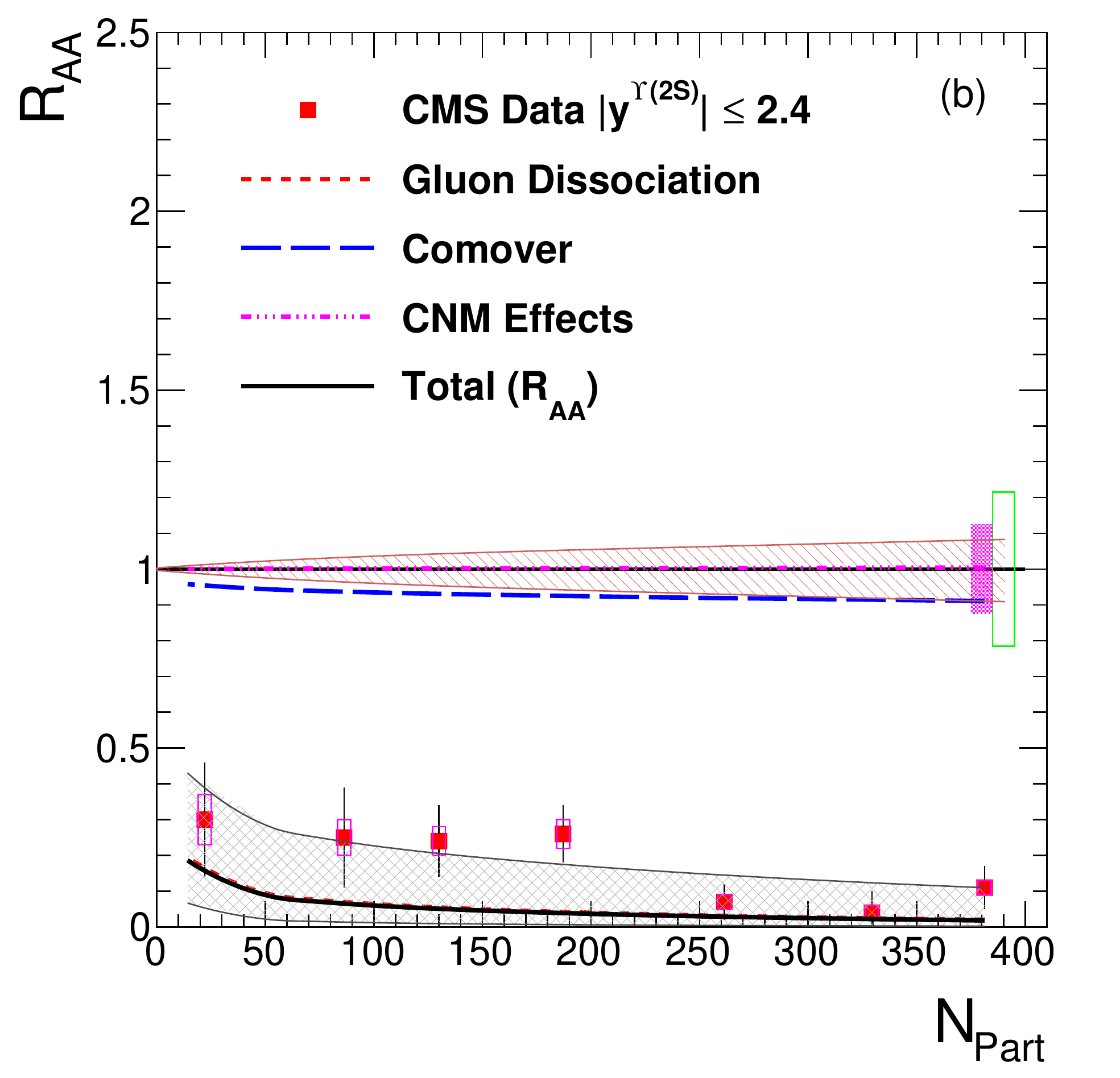}
\caption{(Color online) Calculated nuclear modification factor ($R_{AA}$) compared with CMS 
(a) $\Upsilon$(1S) and (b) $\Upsilon$(2S) measurements. Regeneration is assumed to be negligible. }
\label{fig:UpsilonRaa}
\end{figure}
\begin{figure}
\includegraphics[width=0.49\textwidth]{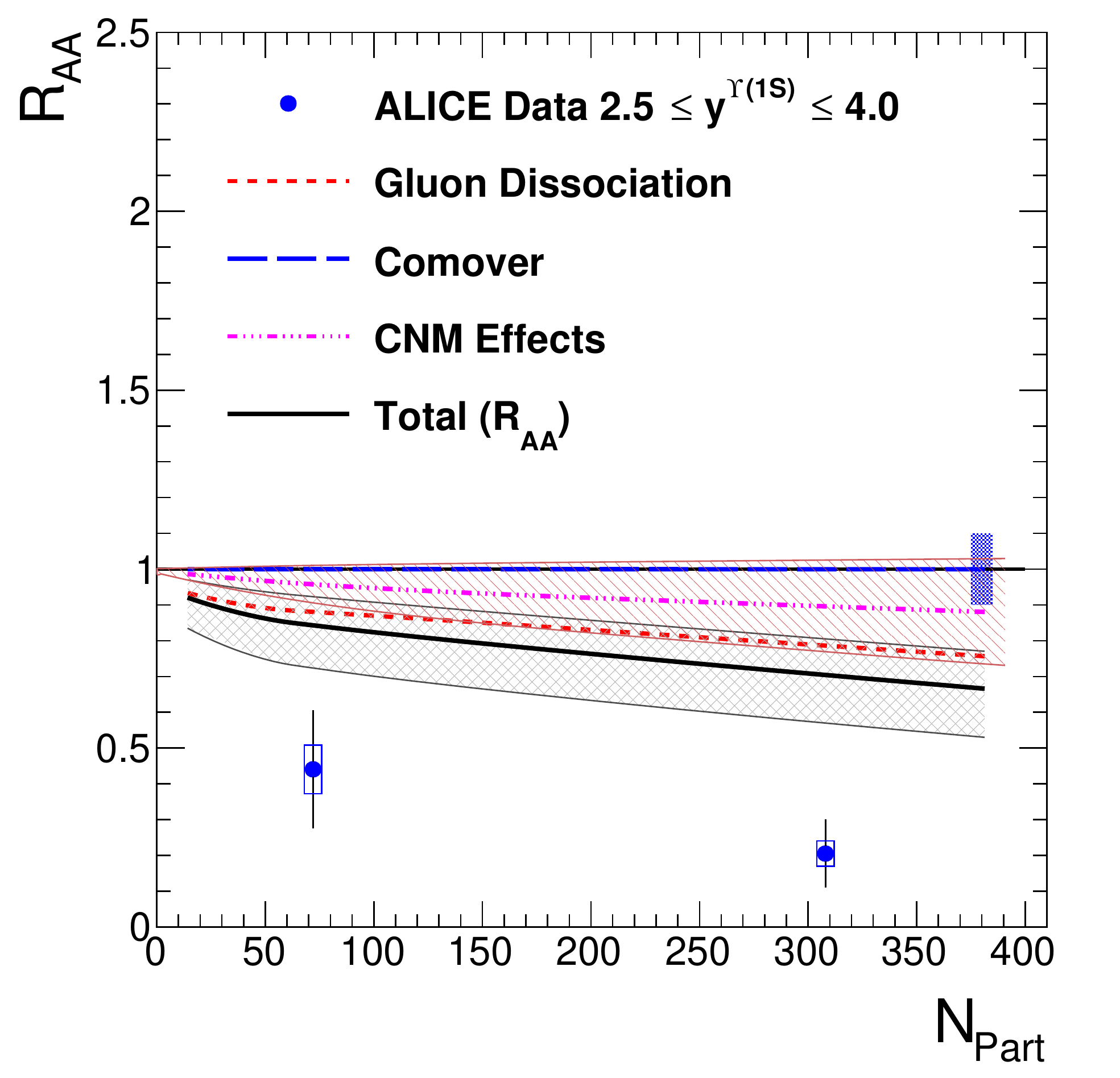}
\caption{(Color online) Calculated nuclear modification factor ($R_{AA}$) compared with 
  ALICE $\Upsilon$(1S) measurement in forward rapidity.}
\label{fig:ALICERaaY}
\end{figure}
\section{hadronic comovers}
  The suppression of quarkonia by comoving pions can be calculated by folding the quarkonium-pion
dissociation cross section $\sigma_{\pi Q}$ over thermal pion distributions \cite{Vogt:1988fj}. 
It is expected  that at LHC energies, the comover cross section will be small~\cite{Lourenco:2008sk}.
{\color{black}
The pion-quarkonia cross section is calculated by convoluting the gluon-quarkonia cross section $\sigma_D$
over the gluon distribution inside the pion~\cite{Arleo:2001mp},
\begin{equation}
\sigma_{\pi Q} (p_{\pi}) = {p_+^2 \over 2(p_\pi^2 - m_\pi^2)} \int_0^1 \, dx \, G(x) \, \sigma_D(xp_+/\sqrt {2}),
\end{equation}
where $p_+ = (p_\pi + \sqrt{p_\pi^2-m_\pi^2})/\sqrt{2}$. The gluon distribution, $G(x)$, inside a pion is 
given by the GRV parameterization~\cite{Gluck:1991ey}. 
The pion momentum $p_\pi$ is related to center of mass energy $\sqrt{s}$ of pion-$J/\psi$ system by 
$p_\pi = (s-M_Q^2-m_\pi^2)/(2M_Q)$.}
The dissociation rate $\lambda_{D_{\pi}}$  can be written as
\begin{eqnarray}
  \lambda_{D_{\pi}} \, \rho_{\pi} & = & \frac{g_\pi}{(2\pi)^{3}} \int d^{3}p_{\pi} f_{\pi}(p) \sigma_{\pi Q} (s) v_{\rm rel} (s) \\ \nonumber
                              & = &\frac{g_\pi}{(2\pi)^{3}} \int\,dp_{\pi}\,2\pi p_{\pi}^{2} f_{\pi}(p_{\pi}) \int\,d{\rm cos}\theta\,\sigma_{\pi Q}(s) \, v_{\rm rel}(s) \Theta(s-4m_{D}^{2}),  \\\nonumber
\end{eqnarray}
where $f_{\pi}(p_{\pi},T)$ is the thermal pion distribution. The  pion density $\rho_{\pi}$ is 
\begin{eqnarray}
\rho_\pi =\frac{g_\pi}{(2\pi)^{3}} \int d^3p_{\pi} \, f_{\pi}(p_{\pi}). 
\end{eqnarray}
The survival probability from pion collisions at freeze-out time $\tau_f$ is written as
\begin{equation}
S_\pi(p_T) = \exp \left( {-\int_{\tau_0}^{\tau_f} \,d\tau\,(1-f(\tau)) \lambda_{D_{\pi}}(T,p_T)\,\rho_{\pi}(T)} \right).
\end{equation}
The hadronic fraction (1-$f(\tau)$) is zero in QGP phase.
The probability $S_\pi(p_T)$ multiplies $S(p_T)$ in Eq.~(\ref{raa}).

\section{Results and discussion}
Figure~\ref{fig:JPsiRaaVsPt}(a) shows the contributions to the nuclear modification factor,
$R_{AA}$, for the $\Jpsi$ as a function of $p_T$ compared with ALICE measurements~\cite{Abelev:2013ila}. 
Figure~\ref{fig:JPsiRaaVsPt}(b) shows the same for the CMS high $p_T$ measurements~\cite{Mironov:2013jaa}. 
At low $p_T$, regeneration of $\Jpsi$ is the dominant process and this seems to be the 
reason for the enhancement of $\Jpsi$ in the ALICE low $p_T$ data.
The gluon suppression is also substantial at low $p_T$ and reduces as we move to high $p_T$. 
Both of these processes (regeneration and dissociation) due to the presence of QGP are 
at play at low and intermediate $p_T$. The high $p_T$ suppression ($p_T > 10$  GeV/$c$) of $\Jpsi$ 
measured by CMS is greater than that due to dissociation by gluons in the QGP. {\color{black} We note 
that at the highest $p_T$ values from CMS, $p_T \gg M_Q$, and energy loss might play a 
similar role for the $J/\psi$ at this $p_T$ as it does for open charm. So the large suppression observed in the high $p_T$
region may be due to energy loss inside the QGP.}
{\color{black} The dominant sources of the uncertainties come from the gluon-quarkonia 
cross section ($\sigma_{D}$) and initial temperature $T_{0}$. We vary the quarkonium-gluon 
cross section by $\pm$ 50\% around the calculated value to obtain the variation in 
the final $R_{AA}$ calculations. The initial temperature is obtained using measured charged particle density and 
assuming $\tau_0$ 0.3 fm/$c$. We vary $\tau_0$ in the range 0.1 $<\,\tau_0\,<$ 0.6 fm/$c$ to 
quantify the uncertainty in $R_{AA}$ corresponding to variation in the initial temperature from 
+45 \% to - 20 \%. Both of these uncertainties are added in quadrature to obtain the final uncertainty band 
around the central value. {\color{black}  The variation of $\tau_0$ and $\sigma_{D}$ results in bands 
on the gluon dissociation and formation curves.  At low $p_T$ the uncertainty in the total 
$R_{AA}$ is driven by the formation while at higher $p_T$, when gluon dissociation is dominant, 
the uncertainty reflects that component.} The uncertainty in the CNM effect is not included in the
$R_{AA}$ uncertainty band since the CNM effects are not dominant.}

We have also calculated $R_{AA}$ as a function of collision centrality (system size).
Figure~\ref{fig:ALICEJPsiRaa_Forward} shows different contributions to the $\Jpsi$ 
nuclear modification factor as a function of system size, along with the ALICE forward rapidity 
measurements~\cite{Abelev:2013ila}. Figure~\ref{fig:ALICEJPsiRaa_Forward} 
indicates that $\Jpsi$'s are increasingly suppressed by the QGP when the system size grows. Since the number 
of regenerated $\Jpsi$'s also grows, the nuclear modification factor remains flat for most of 
the centrality range.
{\color{black}
Figure~\ref{fig:JPsiRaa_Mid} (a) shows the $\Jpsi$ nuclear modification factor along with the ALICE measurement
at midrapidity~\cite{Abelev:2013ila}. Similar to forward rapidity, the nuclear modification factor is
flat in the measured range of $N_{\rm part}$ due to the competitive effects of 
gluon dissociation and regeneration. Our calculations reproduce the measured data within 
uncertainty.} Figure~\ref{fig:JPsiRaa_Mid} (b) 
shows the same for $p_{T}\,\geq$ 6.5 GeV/c, measured by CMS experiment \cite{Mironov:2013jaa}. 
The CMS centrality dependence of the $\Jpsi~R_{AA}$ is well described by the model.
Most of the contribution to the CMS data comes from $\Jpsi$'s with 6.5 $<\,p_T<\,$10 GeV/$c$ where 
the suppression is predominantly due to gluon dissociation.

Figure~\ref{fig:UpsilonRaa} (a) demonstrates the contributions from different processes to the 
centrality dependence of the $\Upsilon$(1S) nuclear modification factor, along with the midrapidity 
data from CMS~\cite{Chatrchyan:2012lxa}. The calculations underestimate the suppression but reproduce 
the shape of centrality dependence. This may be due to the feed down effects from the excited states. 
Figure~\ref{fig:UpsilonRaa} (b) shows the same for the $\Upsilon$(2S) nuclear modification factor
along with the CMS measurements at midrapidity. The excited $\Upsilon$(2S) states 
are highly suppressed. The effect of regeneration, not shown, is negligible 
for the $\Upsilon$ states. 
{\color{black} Figure~\ref{fig:ALICERaaY} shows the forward rapidity 
  ALICE measurement of the $\Upsilon$(1S) nuclear modification factor \cite{Abelev:2014nua}
  along with our calculations. The suppression due to thermal gluon dissociation is smaller 
  than the measured suppression which may be due to the effect of feed down from the $\Upsilon$(2S)
  and higher states.} 
{\color{black} However the measurement is consistent with the suppression of $\Upsilon$(2S) and  
  $\Upsilon$(3S) contribution, along with suppression of the $\Upsilon$(1S) by gluon 
  dissociation.
}
\section{Summary}
 We have carried out detailed calculations of the $\Jpsi$ and $\Upsilon$ 
 modifications in Pb+Pb collisions at LHC.
 The quarkonia and heavy flavour cross sections calculated up to NLO are used in the study. 
 Shadowing corrections are obtained with the EPS09 NLO parametrization.
 A kinetic model is employed which incorporates quarkonia suppression inside QGP, suppression 
 due to hadronic comovers and regeneration from charm pairs.
 The dissociation and formation rates have been studied as a function of medium temperature
 and transverse momentum.
 The nuclear modification factors for $\Jpsi$ and $\Upsilon$  as a function of centrality and 
 transverse momentum have been compared to the measurements in 
 Pb+Pb collisions at $\sqrt{s_{_{NN}}}$ =  2.76 TeV.
 At low $p_T$, regeneration of $\Jpsi$ is the dominant process and this seems to be the process
 for the enhancement of $\Jpsi$ in the ALICE low $p_T$ data.
 Gluon dissociation is also substantial at low $p_T$ and becomes small as we move to high $p_T$. 
 Both of these processes (regeneration and dissociation) due to the presence of QGP affec the 
 yields of quarkonia at low and intermediate $p_T$. The high $p_T$ 
 suppression ($p_T > 10$  GeV/$c$) of $\Jpsi$ measured by CMS is far more than expected due 
 to the dissociation by gluons in QGP.
 The centrality dependence of nuclear modification indicates that  $\Jpsi$'s are increasingly 
 suppressed  when system size grows. Since the number of regenerated  $\Jpsi$'s also grows, the nuclear 
 modification factor of low $p_T$ measurements (ALICE case) remains flat for most of 
 the centrality region. 
 The centrality dependence of $R_{AA}$ for high $p_T$ $\Jpsi$'s is also well described by the model.
 The centrality dependence of suppression of $\Upsilon$ states are reproduced
 by model calculations. Feed down corrections seems to be important for $\Upsilon$(1S).

 \section{Acknowledgement}
  The authors thank their CMS colleagues for the fruitful discussions, 
help and comments. Many of these results were presented at WHEPP and we acknowledge discussions 
with the participants of the meeting, in particular with D. Das, S. Datta, R. Gavai, S. Gupta
and R. Sharma. The work of RV was performed under the auspices of the US Department of Energy, 
Lawrence Livermore National Laboratory, Contract DE-AC52-07NA27344.


\noindent
\begin{thebibliography}{100}
\medskip

\bibitem{Matsui:1986dk} 
 T.~Matsui and H.~Satz,
 ``$J/\psi$ Suppression by Quark-Gluon Plasma Formation'',
 Phys.\ Lett.\ B {\bf 178}, 416 (1986).

\bibitem{Schukraft:2013wba} 
  J.~Schukraft,
  ``Heavy Ion Physics at the LHC: What's new ? What's next ?'',
  arXiv:1311.1429 [hep-ex].


\bibitem{Kluberg:2009wc} 
  L.~Kluberg and H.~Satz,
  ``Color Deconfinement and Charmonium Production in Nuclear Collisions,''
  arXiv:0901.3831 [hep-ph].

\bibitem{Brambilla:2010cs} 
  N.~Brambilla, S.~Eidelman, B.~K.~Heltsley, R.~Vogt, G.~T.~Bodwin, E.~Eichten, A.~D.~Frawley and A.~B.~Meyer {\it et al.},
  ``Heavy quarkonium: progress, puzzles, and opportunities,''
  Eur.\ Phys.\ J.\ C {\bf 71}, 1534 (2011).

\bibitem{Adare:2011yf} 
  A.~Adare {\it et al.}  [PHENIX Collaboration],
  ``$J/\psi$ suppression at forward rapidity in Au+Au collisions at $\sqrt{s_{NN}}=200$ GeV,''
  Phys.\ Rev.\ C {\bf 84}, 054912 (2011).

\bibitem{Andronic:2003zv} 
  A.~Andronic, P.~Braun-Munzinger, K.~Redlich and J.~Stachel,
  ``Statistical hadronization of charm in heavy ion collisions at SPS, RHIC and LHC,''
  Phys.\ Lett.\ B {\bf 571}, 36 (2003).

\bibitem{Muller:2012zq} 
  B.~Muller, J.~Schukraft and B.~Wyslouch,
  ``First Results from Pb+Pb collisions at the LHC,''
  Ann.\ Rev.\ Nucl.\ Part.\ Sci.\  {\bf 62}, 361 (2012).

\bibitem{P.ShuklaforCMS:2014vna} 
  P.~Shukla [CMS Collaboration],
  ``Overview of quarkonia and heavy flavour measurements by CMS,''
  arXiv:1405.3810 [nucl-ex].

\bibitem{Chatrchyan:2012np} 
  S.~Chatrchyan {\it et al.}  [CMS Collaboration],
  ``Suppression of non-prompt $J/\psi$, prompt $J/\psi$, and Y(1S) in Pb+Pb collisions at $\sqrt{s_{NN}}=2.76$ TeV,''
  JHEP {\bf 1205}, 063 (2012).

\bibitem{Mironov:2013jaa} 
  C.~Mironov [CMS Collaboration],
  ``Overview of results on heavy flavour and quarkonia from the CMS Collaboration,''
  Nucl.\ Phys.\ A {\bf 904-905}, 194c (2013).

\bibitem{Tang:2011kr} 
  Z.~Tang [STAR Collaboration],
  ``$J/\psi$ production and correlation in p+p and Au+Au collisions at STAR,''
  J.\ Phys.\ G {\bf 38}, 124107 (2011).

\bibitem{Abelev:2013ila} 
  B.~B.~Abelev {\it et al.}  [ALICE Collaboration],
  ``Centrality, rapidity and transverse momentum dependence of J/$\psi$ suppression in Pb-Pb collisions at $\sqrt{s_{NN}}$=2.76 TeV,''
  Phys.\ Lett.\  {\bf 743}, 314 (2014).

\bibitem{Chatrchyan:2011pe} 
  S.~Chatrchyan {\it et al.}  [CMS Collaboration],
  ``Indications of suppression of excited $\Upsilon$ states in PbPb collisions at $\sqrt{S_{NN}}$ = 2.76 TeV,''
  Phys.\ Rev.\ Lett.\  {\bf 107}, 052302 (2011).

\bibitem{Chatrchyan:2012lxa} 
  S.~Chatrchyan {\it et al.}  [CMS Collaboration],
  ``Observation of sequential Upsilon suppression in PbPb collisions,''
  Phys.\ Rev.\ Lett.\  {\bf 109}, 222301 (2012).


 \bibitem{Abelev:2014nua} 
  B.~B.~Abelev {\it et al.}  [ALICE Collaboration],
  ``Suppression of $\Upsilon$(1S) at forward rapidity in Pb-Pb collisions at $\sqrt{s_{\rm NN}}$ = 2.76 TeV,''
  arXiv:1405.4493 [nucl-ex].


\bibitem{Abdulsalam:2012bw} 
  A.~Abdulsalam and P.~Shukla,
  ``Suppression of bottomonia states in finite size quark gluon plasma in PbPb collisions at Large Hadron Collider,''
  Int.\ J.\ Mod.\ Phys.\ A {\bf 28}, 1350105 (2013).

\bibitem{Bhanot:1979vb} 
  G.~Bhanot and M.~E.~Peskin,
  ``Short Distance Analysis for Heavy Quark Systems. 2. Applications,''
  Nucl.\ Phys.\ B {\bf 156}, 391 (1979).

\bibitem{Xu:1995eb} 
  X.~-M.~Xu, D.~Kharzeev, H.~Satz and X.~-N.~Wang,
  ``J/$\psi$ suppression in an equilibrating parton plasma,''
  Phys.\ Rev.\ C {\bf 53}, 3051 (1996).

\bibitem{Andronic:2012dm} 
  A.~Andronic, P.~Braun-Munzinger, K.~Redlich and J.~Stachel,
  ``The statistical model in Pb-Pb collisions at the LHC,''
  Nucl.\ Phys.\ A {\bf 904-905}, 535c (2013).

\bibitem{Thews:2000rj} 
  R.~L.~Thews, M.~Schroedter and J.~Rafelski,
  ``Enhanced $J/\psi$ production in deconfined quark matter,''
  Phys.\ Rev.\ C {\bf 63}, 054905 (2001).

\bibitem{Vogt:2010aa} 
  R.~Vogt,
  ``Cold Nuclear Matter Effects on $J/\psi$ and $\Upsilon$ Production at the LHC,''
  Phys.\ Rev.\ C {\bf 81}, 044903 (2010).

\bibitem{Zhao:2011cv} 
  X.~Zhao and R.~Rapp,
  ``Medium Modifications and Production of Charmonia at LHC,''
  Nucl.\ Phys.\ A {\bf 859}, 114 (2011).

\bibitem{Emerick:2011xu} 
  A.~Emerick, X.~Zhao and R.~Rapp,
  ``Bottomonia in the Quark-Gluon Plasma and their Production at RHIC and LHC,''
  Eur.\ Phys.\ J.\ A {\bf 48}, 72 (2012).

\bibitem{Strickland:2011mw} 
  M.~Strickland,
  ``Thermal $\Upsilon$(1S) and $\chi_{b1}$ suppression in $\sqrt{s_{NN}}=2.76$ TeV Pb-Pb collisions at the LHC,''
  Phys.\ Rev.\ Lett.\  {\bf 107}, 132301 (2011).



\bibitem{Lai:2010vv} 
  H.~L.~Lai, M.~Guzzi, J.~Huston, Z.~Li, P.~M.~Nadolsky, J.~Pumplin and C.-P.~Yuan,
  ``New parton distributions for collider physics,''
  Phys.\ Rev.\ D {\bf 82}, 074024 (2010).






\bibitem{Nelson:2012bc}
  R.~E.~Nelson, R.~Vogt and A.~D.~Frawley,
  `Narrowing the uncertainty on the total charm cross section and its effect on the J/$\psi$ cross section,''
  Phys.\ Rev.\ C {\bf 87}, no. 1, 014908 (2013).

\bibitem{Nelson:Future}
  R. Nelson, R. Vogt and A. D. Frawley, in preparation.


\bibitem{Cacciari:2005rk} 
  M.~Cacciari, P.~Nason and R.~Vogt,
  ``QCD predictions for charm and bottom production at RHIC,''
  Phys.\ Rev.\ Lett.\  {\bf 95}, 122001 (2005).


\bibitem{Eskola:2009uj} 
  K.~J.~Eskola, H.~Paukkunen and C.~A.~Salgado,
  ``EPS09: A New Generation of NLO and LO Nuclear Parton Distribution Functions,''
  JHEP {\bf 0904}, 065 (2009).




\bibitem{Kumar:2012qx} 
  V.~Kumar, P.~Shukla and R.~Vogt,
  ``Components of the dilepton continuum in Pb+Pb collisions at $\sqrt{s_{_{NN}}} = 2.76 $ TeV,''
  Phys.\ Rev.\ C {\bf 86}, 054907 (2012).

\bibitem{Chatrchyan:2011sx} 
  S.~Chatrchyan {\it et al.}  [CMS Collaboration],
  ``Observation and studies of jet quenching in PbPb collisions at nucleon-nucleon center-of-mass energy = 2.76 TeV,''
  Phys.\ Rev.\ C {\bf 84}, 024906 (2011).

\bibitem{Huovinen:2009yb} 
  P.~Huovinen and P.~Petreczky,
  ``QCD Equation of State and Hadron Resonance Gas,''
  Nucl.\ Phys.\ A {\bf 837}, 26 (2010).
  
\bibitem{Aamodt:2010cz} 
  K.~Aamodt {\it et al.}  [ALICE Collaboration],
  ``Centrality dependence of the charged-particle multiplicity density at mid-rapidity in Pb-Pb collisions at $\sqrt{s_{NN}}=2.76$ TeV,''
  Phys.\ Rev.\ Lett.\  {\bf 106}, 032301 (2011).
  

\bibitem{Chatrchyan:2011pb} 
  S.~Chatrchyan {\it et al.}  [CMS Collaboration],
  ``Dependence on pseudorapidity and centrality of charged hadron production in PbPb collisions at a nucleon-nucleon 
  centre-of-mass energy of 2.76 TeV,''
  JHEP {\bf 1108}, 141 (2011).
  
\bibitem{Shuryak:1992wc} 
  E.~V.~Shuryak,
  ``Two stage equilibration in high-energy heavy ion collisions,''
  Phys.\ Rev.\ Lett.\  {\bf 68}, 3270 (1992).



\bibitem{Abbas:2013bpa} 
  E.~Abbas {\it et al.}  [ALICE Collaboration],
  ``Centrality dependence of the pseudorapidity density distribution for charged particles 
  in Pb-Pb collisions at $\sqrt{s_{\rm NN}}$ = 2.76 TeV,''
  Phys.\ Lett.\ B {\bf 726}, 610 (2013).
  
  
\bibitem{Karsch:1987pv} 
  F.~Karsch, M.~T.~Mehr and H.~Satz,
  ``Color Screening and Deconfinement for Bound States of Heavy Quarks,''
  Z.\ Phys.\ C {\bf 37}, 617 (1988).
  
\bibitem{Arleo:2001mp} 
  F.~Arleo, P.~B.~Gossiaux, T.~Gousset and J.~Aichelin,
  ``Heavy quarkonium hadron cross section in QCD at leading twist,''
  Phys.\ Rev.\ D {\bf 65}, 014005 (2002).
  
\bibitem{Thews:2005vj} 
  R.~L.~Thews and M.~L.~Mangano,
  ``Momentum spectra of charmonium produced in a quark-gluon plasma,''
  Phys.\ Rev.\ C {\bf 73}, 014904 (2006).
  
\bibitem{Vogt:1988fj} 
  R.~Vogt, M.~Prakash, P.~Koch and T.~H.~Hansson,
  ``$J/\psi$ Interactions With Hot Hadronic Matter,''
  Phys.\ Lett.\ B {\bf 207}, 263 (1988).
  
\bibitem{Lourenco:2008sk} 
  C.~Lourenco, R.~Vogt and H.~K.~Woehri,
  ``Energy dependence of J/$\psi$ absorption in proton-nucleus collisions,''
  JHEP {\bf 0902}, 014 (2009).
  

\bibitem{Gluck:1991ey} 
  M.~Glueck, E.~Reya and A.~Vogt,
  ``Pionic parton distributions,''
  Z.\ Phys.\ C {\bf 53}, 651 (1992).
  

  

\end{thebibliography}
\end{document}